\documentclass[nofootinbib,showpacs,preprintnumbers,amsmath,amssymb,floatfix]
{revtex4-2}
\usepackage{graphicx}

\begin{document}

\title{Proton spin in double-logarithmic approximation}

\author{B. I. Ermolaev}

\affiliation{Ioffe Institute, 194021
 St.Petersburg, Russia}


\begin{abstract}
Proton is a composite particle, so its spin $S_P$ is made from the spins of the partons which the proton consists of.
 The discrepancy between $S_P= 1/2$ and the experimentally detected sum of the parton spins
was named Proton Spin Puzzle.
Solution to this problem includes formulae for the parton helicities valid in
the whole range of $x$. There are approaches in the literature for calculating the helicities.
As  a theoretical basis they apply evolution equations of different types.
Despite these equations are constructed for operating in widely different regions of $x$ and account for different contributions,
all of them  equally well suited for solving the proton spin problem.
Our explanation of  this situation is that the main impact on values of the parton spin contributions
should be brought not by the evolution equations themselves but by  phenomenological
  fits for initial parton distributions.

 We suggest a more theoretically grounded  approach to description of the parton helicities and apply it to
 solving the proton spin problem. It combines the total resummation of double logarithms (DL),
 accounting for the running $\alpha_s$ effects and DGLAP
formulae, leading to expressions for the helicities valid at arbitrary $x$. As a consequence, the set of involved
phenomenological parameters in our approach is minimal and its influence on the helicity behaviour is weak.
 We apply our approach to solve the proton spin problem in the straightforward
 way and make an estimate,
demonstrating that the RHIC data complemented by the DL contributions from
the regions of $x$ beyond the RHIC scope
are well compatible with the value $S_P = 1/2$.
\end{abstract}


\maketitle

\section{Introduction}
\label{intro}

The proton spin puzzle was first reported in  EMC publications Refs.~\cite{emc1,emc2} and then continued in Refs.~\cite{spin1} - \cite{spin9}. The point is
that the experimental data on the spin content of quarks and gluons in the proton
disagreed with the obvious requirement of the
total angular moment conservation.
Indeed, the angular moment conservation prescribed that\footnote{notice that Orbital Angular Momentum contributions were dropped in Refs.~\cite{emc1}-\cite{spin9}}
\begin{equation}\label{spdef}
S_P = S_q + S_g =1/2,
\end{equation}
while EMC reported that

\begin{equation}\label{spdefemc}
 S_q + S_g  < 1/2.
\end{equation}

In Eq.~(\ref{spdef},\ref{spdefemc}), $S_P$ is the proton spin and $S_q, S_g$ are the quark and gluon spin contributions respectively.
The contradiction between Eqs.(\ref{spdefemc}) and (\ref{spdef}) was totally unexpected and was named the
proton spin puzzle/spin crisis.
Later, it was suggested in Refs.~\cite{oam1,oam2} to add the
quark and gluon Orbital Angular Momentum (OAM) contributions, $L_{q,g}$ to the spin contributions in Eq.~(\ref{spdef}), converting Eq.~(\ref{spdef}) into

\begin{equation}\label{spoam}
S_P = S_q + S_g + L_q + L_g.
\end{equation}
Eq.~(\ref{spoam}) has been commonly accepted nowadays.
The parton contributions $S_{q,g}$ are expressed through the helicity distributions $\Delta \Sigma (x)$ (for quarks) and $\Delta G (x)$ (for gluons):

\begin{eqnarray}\label{sqsg}
S_q &=& \frac{1}{2} \int_0^1 dx \Delta \Sigma (x),
\\ \nonumber
 S_g &=&  \int_0^1 dx \Delta G(x).
\end{eqnarray}

The recent RHIC data\cite{rhic1,rhic2} are

\begin{eqnarray}\label{sexp}
0.15 \leq \bar{S}_q \leq 0.20,
~~~0.13 \leq \bar{S}_g \leq 0.26,
\end{eqnarray}
where the range of involved values of $x$ is
\begin{equation}\label{x1}
1  \geq x \geq x_1 = 0.001
\end{equation}
for $\bar{S}_q $ and

\begin{equation}\label{x2}
 1 \geq x \geq x_2 = 0.05
\end{equation}
for $\bar{S}_g $.
We name $\Delta \bar{S}_q$ and $\Delta \bar{S}_g$ the intervals of $S_{q,g}$ corresponding to Eq.~(\ref{sexp}).
Values of $S_{q,g}$ from the intervals $\Delta \bar{S}_{q,g}$ are incompatible with Eq.~(\ref{spdef}). It is illustrated in Fig.~\ref{helfig2},
\begin{figure}\label{helfig2}
\includegraphics[width=.6\textwidth]{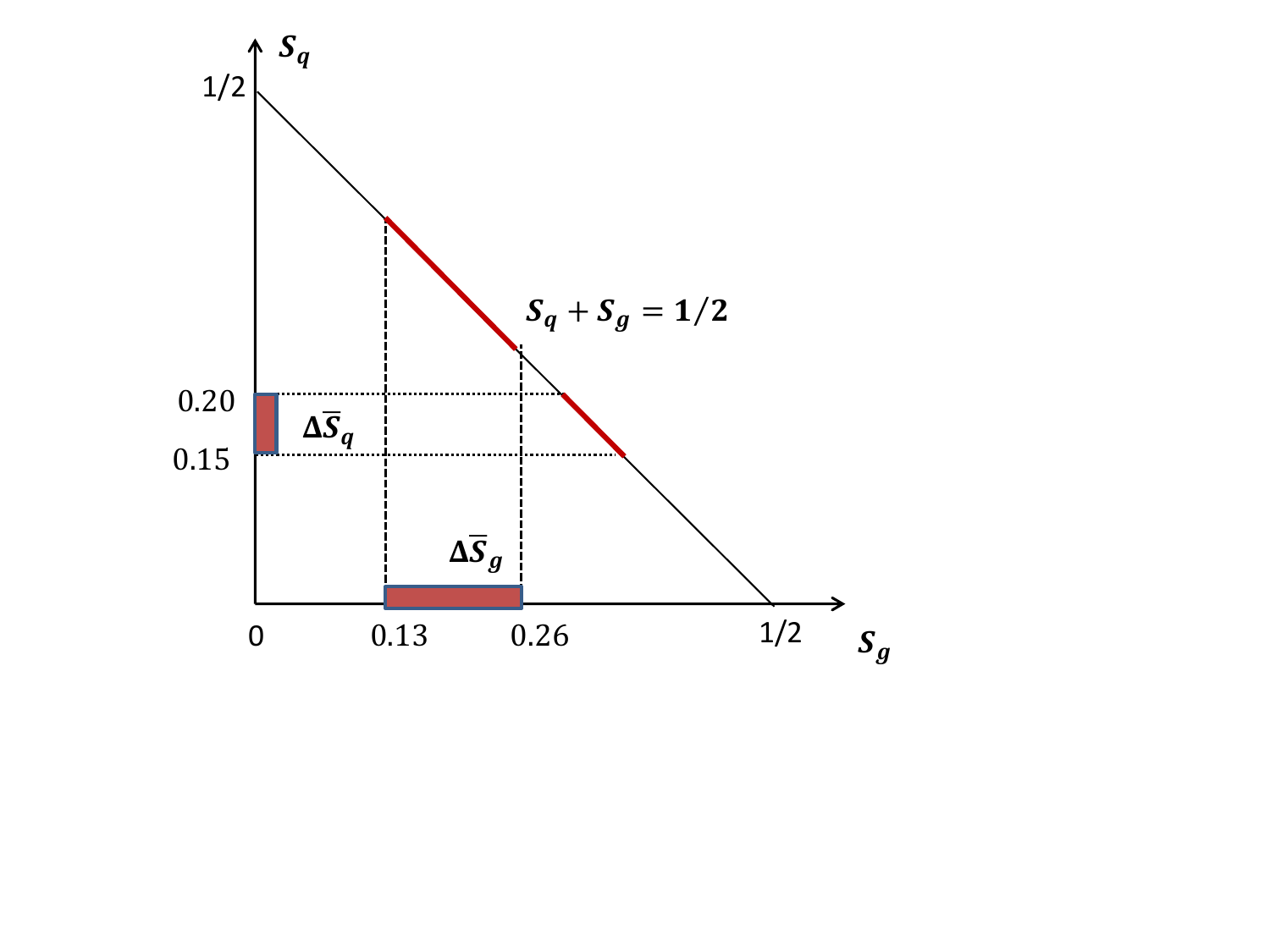}
\caption{\label{helfig2}
Projections of $\Delta \bar{S}_q$ and $\Delta \bar{S}_g$ on the line $\bar{S}_q + \bar{S}_g = 1/2$ do not overlap, i.e. the
values of $S_{q,g}$ from these intervals do not obey the requirement $S_q + S_g = 1/2$.}
\end{figure}
where coordinates of all dots
on the slanted line satisfy the condition $S_q + S_g = 1/2$ but none of them obeys the requirement $[S_q \in \Delta \bar{S}_q]\bigcap [S_g \in \Delta \bar{S}_g]$.
On the other hand, substituting  $\bar{S}_q$ and  $\bar{S}_g$ in Eq.~(\ref{spdef}) instead of $S_q$ and $S_g$
does not mean violation of Eqs.~(\ref{spdef},\ref{spoam}) because it
leaves uninspected the intervals $x < x_1$ for quarks and $x < x_2$ for gluons.

Calculating the double-logarithmic (DL) contributions to $S_{q,g}$ and $L_{q,g}$
was the subject of investigation in Refs.~\cite{agr}
-\cite{global}.
The method of calculations
(KPSCTT)
was originally suggested in Refs.~\cite{kovchfirst}-\cite{kovch3} and
then improved in Refs.~\cite{kovch2}-\cite{smalldis}. Solution to KPSCTT equations were expressed in terms of small-$x$ asymptotics.
To be more precise, there were calculated the $x$-dependent parts of the small-$x$ asymptotics of the parton helicities
for
the particular case when the quark contributions were neglected. One of the results obtained in Refs.~\cite{kovchfirst}-\cite{smalldis} was
 verification and
confirmation of the intercepts of the small-$x$ asymptotics of the structure function $g_1$  calculated earlier
in Refs~\cite{emr,berns,bers}  by other means.
From the theoretical point of view, the most essential flaw of KPSCTT is that this method operates with the small-$x$ asymptotics only
whilst integrating in Eq.~(\ref{sqsg}) involves  values of $x$ outside the  asymptotics applicability region, where
contributions of the asymptotics are small. In order to make them greater, Refs.~\cite{agr}
-\cite{global} use a set of phenomenological parameters. We discuss KPSCTT in more detail in Sect.~IIA.


 In contrast, Refs.~\cite{w1,w2} do not involve asymptotics and operate with the NLO and  NNLO DGLAP  respectively.
However, DGLAP  was originally suggested for operating  at $x \sim 1$. The coefficient functions calculated in DGLAP
do not develop the Regge type of the small-$x$ asymptotics and do not provide  $g_1$ with
 the necessary rise at small $x$.
Such fast growth in the DGLAP framework is ensured by the phenomenological singular factors $ x^{-a}$ (with $a > 0$) installed ad hoc in the initial parton densities. In other words, the Regge asymptotics are introduced into DGLAP technology through the fits.
Such factors mimic resummation of leading (Double-Logarithmic) contributions (see Ref.~\cite{egtg1c} for detail) and should
be dropped when the resummation is accounted for. By construction, DGLAP fits
are organized in the way which increases the impact of not so small $x$. For example, the exponent $a$ is
greater than the genuine intercept. It makes value of  $ x^{-a}$ be great at  pre-asymptotic $x$.
Also note that the usual DGLAP parametrization $\alpha_s = \alpha_s (k^2_{\perp})$ fails at small $x$. These topics are discussed in Sect.~IIB,C.

Despite such a significant  difference between the theoretical apparatus in Refs.~\cite{w1,w2} and \cite{kovch2}-\cite{smalldis}
and despite the fact that none of them can be used in the whole range of $x$,
the both approaches present the results which
 are in a good agreement with $S_P = 1/2$. Explanation of this agreement is that the lack of theoretical
 basis in each of the approaches is compensated by impact of the phenomenological parameters in the fits.

We suggest a more theoretically grounded approach to describe the parton helicities. Its perturbative component is valid in the
whole range of $x$: $0 \leq x \leq 1$, and accounts for the most essential contributions at any $x$. As a result, the set of
phenomenological parameters is significantly reduced.  Namely, we describe the quark and gluon helicities with expressions
obtained in the Double-Logarithmic Approximation (DLA).
  The point is that the DL contributions $\sim \left(\alpha_s \ln^2 \left(1/x\right)\right)^n$, ($n = 1,2,..$), are the leading contributions
 at small $x$, so the total summation of them is important in the small-$x$ kinematics.
In addition, we keep $\alpha_s$ running in each vertex
 of the involved Feynman graphs. This issue is considered in Sect.~IIC.
 Complementing the DL terms by NLO DGLAP contributions
  allows us to obtain the formulae valid at arbitrary $x$. Such expressions were obtained in Refs.~\cite{egtg1a,egtg1b,egtg1c}.
On one hand, they lead to the Regge asymptotic behavior at very small $x$; on the other hand, they coincide with the DGLAP formulae at $x \sim 1$.

As for the technical means,
we use the modification of
the Infra-Red Evolution Equations (IREE) approach\footnote{History, details and application of the IREE method to DIS can be found in
Ref.~\cite{egtg1c}.} suggested by L.N.~Lipatov in Ref.~\cite{l} and applied to the elastic quark scattering in
Ref.~\cite{kl}.  The key idea of the original IREE method is
tracing evolution with respect to the infrared (IR) cut-off $\mu$.
Introducing $\mu$ to regulate IR singularities is necessary because all DL contributions stem from IR singularities, which
is trivial for DL from virtual gluons and
stands for DL from virtual quarks when the quark masses are neglected. The IREE approach makes it possible to
to collect
DL contributions to all orders in $\alpha_s$. Setting the scale for $\mu$ is discussed in Sect.~IID.
The expressions for the parton helicities in Refs.~\cite{egtg1a,egtg1b,egtg1c} were obtained for $Q^2 \sim \mu^2$, so that their dependency
on $Q^2$ was neglected.
On the other hand, the RHIC data of Eq.~(\ref{sexp}) were taken at fixed value of $Q^2= 10$~GeV$^2$ without tracing any $Q^2$-dynamics. It allows us to focus on the $x$-dependence of the parton helicities
as the first step, neglecting their $Q^2$-dependence, especially as our preliminary estimates showed that the impact of the $Q^2$-dependence on our
conclusions was small at $Q^2 \leq 10$~GeV$^2$. So, we focus in the present paper on the $x$-evolution  and
will investigate the impact of the $Q^2$-dynamics in a separate paper.

Our paper is organized as follows: in Sect.~II we briefly
comment on
Refs.~\cite{agr}
-\cite{w2}.  Then we consider parametrization of $\alpha_s$ at small $x$ and compare it with the standard DGLAP parametrization.
After that we discuss the structure function $g_1$ in DLA and the applicability region of the small-$x$
asymptotics.
In Sect.~III we demonstrate how to
calculate $S_{q,g}$ in DLA in the straightforward way.
However, the expressions for the parton helicities in DLA are quite complicated  to use.
Because of that we suggest in Sect.~IV  a shortcut:
we formulate a simple but consistent approximation for the parton helicities. The
expressions obtained there
include free parameters related to non-perturbative contributions. They are specified through the use of the RHIC data.
Combining the obtained expressions for $S_{q,g}$ with the RHIC data, we arrive at
the parton spin contributions satisfying Eq.~(\ref{spdef}). More precisely, we demonstrate in Sect.~IV that
the RHIC data on $S_{q,g}$ in the intervals (\ref{x1},\ref{x2}) being complemented by the contributions from the $x$-regions beyond these intervals
are compatible with the requirement $S_P = 1/2$ even
when OAM has not been taken into account.  Finally, Sect.~V is for concluding remarks.

\section{Comment on the  KPSCTT, DGLAP and IREE approaches}
\label{sectii}

In this Sect. we briefly comment on the approaches  in Refs.~\cite{kovchfirst}-\cite{smalldis}
and \cite{w1,w2}, then  discuss parametrization of $\alpha_s$ at small $x$ and
remind basic facts about $g_1$ in DLA, including the small-$x$ asymptotics.

\subsection{Comment on KPSCTT}
\label{sec:sectiia}

KPSCTT approach was developed in Refs.~\cite{kovchfirst}-\cite{smalldis}. Solutions to the evolution equations for the parton helicities
obtained with this method are represented in the form of the small-$x$ asymptotics. This applies both
the parton helicities and the OAM contributions. Technically, in this regard the method KPSCTT is similar to BFKL, where solution to the BFKL equation is expressed through the series
of the asymptotics.
 The asymptotics are
 represented by the  expressions which perfectly agree with the Regge theory based on such profound concepts as Analyticity and Causality,
and in addition they are convenient to handle. However, they should be used within their applicability region only.
This region  was estimated in Ref.~\cite{egtg1c}. We consider this issue in Sect.~IID
and for now remind that the asymptotics $\bar{g}_1$ of $g_1$ can reliably represent $g_1$ at $x \leq x_0 \approx 10^{-6}$.
Notice that $x_0 \ll x_{1,2}$ of Eqs.~(\ref{x1}, \ref{x2}). So, applying KPSCTT at
 $x \sim x_{1,2}$ is groundless. Another drawback of KPSCTT is that $\alpha_s$ in this approach is treated as a constant and is fixed a posteriori.

\subsection{Comment on  Refs.~\cite{w1,w2}}
\label{sec:sectiib}

Refs.~\cite{w1,w2} operate with the NLO and NNLO DGLAP formulae for the coefficient functions and anomalous dimensions. However,
the coefficient functions
calculated with the LO and NLO accuracy cannot ensure an appropriate (Regge-like) growth of
$g_1$ at small $x$. Indeed, the small-$x$ asymptotics of the perturbative component $\breve{g}_1^{LODGLAP}$ of $g_1$ in LO DGLAP is well-known:

\begin{equation}\label{aslodglap}
\breve{g}_1^{LODGLAP} \sim e^{\sqrt{\alpha_s\ln (1/x)\ln (Q^2/Q^2_0)}},
\end{equation}
where $Q^2_0$ is the starting point the $Q^2$-evolution and  $\alpha_s$  is kept fixed. We also have dropped some numerical
factors in the exponent for simplicity. Obviously,  the asymptotics of
Eq.~(\ref{aslodglap}) grows at $x \to 0$ much slower then the Regge asymptotics, see Eq.~(\ref{as}). Accounting for higher-orders
contributions to LO DGLAP does not
change much this situation: It is easy to show that in the NLO DGLAP case the asymptotics $\breve{g}_1^{NLODGLAP}$ is

\begin{equation}\label{asnlodglap}
\breve{g}_1^{NLODGLAP} \sim e^{\alpha^{1/2}_s\ln^{3/4} (1/x)\ln^{1/4} (Q^2/Q^2_0)},
\end{equation}
which is again far from the Regge behavior. The same is true for NNLO DGLAP, etc.
The fast, Regge-like rise is actually achieved with installing the singular factors $x^{-a}$ in the fits for initial parton densities. The form of such fits used in
Refs.~\cite{w1,w2} looks like

\begin{equation}\label{fit}
\Phi (x, Q^2_0) = N x^{- a} (1 - x)^b (1 + c x^d),
\end{equation}
with $Q^2_0 = 1$~GeV$^2$ and $a,b,c,d$  being phenomenological parameters, all of them are positive. We proved (see Ref.~\cite{egtg1c}) that the role
of the factor $x^{-a}$ is mimicking the total resummation of DL contributions. It provides $g_1$ with
the Regge asymptotics: $g_1 \sim x^{-a}$, so $a$ plays the role of the intercept.  When the resummation is taken into account this factor becomes surplus and
should be dropped, which simplifies the fit. Then, the other terms  $\sim x$ in the fit can be dropped at small $x$. In order to simplify the fits of
type of Eq.~(\ref{fit})
at moderate $x$, the DL contributions in the coefficient functions and anomalous dimensions  should be complemented by the NLO DGLAP terms as shown in Ref.~\cite{egtg1c}.
This procedure simplifies Eq.~(\ref{fit}) down to the normalization constant $N$.  We used such inputs in Ref.~\cite{egtcomp}
to explain behavior of $g_1$ at the COMPASS experiments.

\subsection{Treatment of $\alpha_s$ at small $x$}
\label{sec:sectiic}

The QCD coupling $\alpha_s$ in each vertex of every involved Feynman graph should be running. The standard DGLAP parametrization of
$\alpha_s$ in every ladder rung is well-known:

\begin{equation}\label{adglap}
\alpha_s = \alpha_s (k^2_{\perp}),
\end{equation}
with $k_{\perp}$ being the transverse component of the ladder parton. This parametrization leads to the also well-known parametrization
$\alpha_s = \alpha_s (Q^2)$
in the DGLAP equations. There is a factorization between dynamics in the transverse and longitudinal spaces at $x \sim 1$, so the parametrization (\ref{adglap})
of $\alpha_s (k^2_{\perp})$
does not involves longitudinal component of $k$. However, this fails at small $x$, where the factorization between the longitudinal and transverse spaces does not
exists. Because of that $\alpha_s$ related to the space-like gluons is

\begin{equation}\label{asmallx}
\alpha_s \approx \alpha_s (k^2_{\perp}/\beta),
\end{equation}
with $\beta$ being the longitudinal fraction of the same ladder parton momentum $k$.   Obviously, Eq.~(\ref{asmallx})
coincides with the standard DGLAP expression of Eq.~(\ref{adglap}) at $\beta \sim 1$, i.e. in the DGLAP applicability region. Let us remind that integration
 over $k_{\perp}$ at small $x$ does not involve $Q^2$ as the upper limit of integration in contrast to DGLAP.
 In the $\omega$- space, $\alpha_s$ is replaced by $A^{\prime}(\omega)$:

\begin{equation}\label{aprime}
A^{\prime}(\omega) = \frac{1}{b} \Big[\frac{1}{\eta}
- \int_0^{\infty} \frac{d \rho e^{-\omega \rho}}{(\rho + \eta)^2} \Big],
\end{equation}
where $\eta = \ln \left(\mu^2/\Lambda^2_{QCD}\right)$ and $\mu$ is the IR cut-off, see Refs.~\cite{egtalpha,etalpha} for detail.
 When the gluons
are time-like, $\alpha_s$ contains $\pi^2$-terms. As is shown in Ref.~\cite{etalpha}, accounting for them converts Eq.~(\ref{asmallx}) into the coupling $\alpha_s^{eff}$:

\begin{equation}\label{asmallxpi}
\alpha^{eff}_s \approx \alpha_s \left(\mu^2\right) +  \frac{1}{\pi b}\left[\arctan \left(\frac{\pi}{\ln \left(k^2_{\perp}/\beta \Lambda^2_{QCD}\right)}\right)
- \arctan \left(\frac{\pi}{\ln \left(\mu^2/ \Lambda^2_{QCD}\right)}\right)\right]
\end{equation}
and in the $\omega$- space, $\alpha^{eff}_s$ is replaced by $A(\omega)$:

\begin{equation}\label{a}
A(\omega) = \frac{1}{b} \Big[\frac{\eta}{\eta^2 + \pi^2}
- \int_0^{\infty} \frac{d \rho e^{-\omega \rho}}{(\rho + \eta)^2 +
\pi^2} \Big].
\end{equation}

In Eqs.~(\ref{aprime}-
\ref{a}), $b$ is the first coefficient of the Gell-Mann- Low function
and $\mu$ corresponds to beginning of the evolution. Neglecting the $\pi^2$-contribution can be done
when the arguments of the both arctangents are small. Expanding the arctangents  in the power series and retaining the
first terms of the expansions, we are back to Eq.~(\ref{asmallx}). Setting the scale for $\mu$ is discussed in the next Sect.

\subsection{Remark on $g_1$ in DLA}
\label{sec:sectiid}

There is a significant difference between expressions for the structure function $g_1$  and  the parton helicities. Indeed,
Eq.~(\ref{fhsol}) demonstrates that $g_1$ includes  non-linear combinations of
the perturbative components of the helicities. Nevertheless,
 the small-$x$ asymptotics of $g_1$ and the one of
the parton helicities coincide save unessential numerical factors
as is proved in Appendix A. Note that this circumstance was used in Refs.~\cite{kovchfirst}-\cite{smalldis},
albeit without proof. Below we remind some useful features of $g_1$, which can also be applied to the parton helicities.
For the sake of simplicity we consider
first the non-singlet component $g_1^{NS}$ and then discuss the singlet component. It is convenient to represent $g_1^{NS}$
in terms of the Mellin transform (see Ref.~\cite{berns}):

\begin{equation}\label{melling1}
g_1^{NS} (x, Q^2) = \frac{e^2_q}{2} \int_{- \imath \infty}^{\imath
\infty} \frac{d \omega}{2 \pi \imath} x^{- \omega}\; C_{NS}
(\omega) \;e^{h_{NS}(\omega) \ln(Q^2/\mu^2)}\; \Phi (\omega),
\end{equation}
where  $C_{NS}$ is the coefficient function and $h_{NS}$ is the anomalous dimension. They are given by the following expressions:

\begin{equation}\label{chns}
C_{NS} = \frac{\omega}{\omega - h_{NS}(\omega)},~~
h_{NS} = \frac{1}{2} \left[\omega - \sqrt{\omega^2 - \frac{2 \alpha_s C_F}{\pi}\left[1 - \frac{f^{(+)} (\omega)}{2 \pi^2 \omega}\right] }\right],
\end{equation}
with $C_F = 4/3$. The notation $f^{(+)} (\omega)$ stands for the DL color octet contribution to the positive signature scattering amplitude of the forward quark-antiquark annihilation . It was calculated in Ref.~\cite{kl}. Notation $\Phi$ in Eq.~(\ref{melling1}) is for the initial quark distribution and $\mu$ is the infrared cut-off
associated often with the factorization scale.

Although the DGLAP expression for $g_1^{NS}$ has the same form as Eq.~(\ref{melling1}), they differ a lot:
both $C_{NS}$ and $h_{NS}$ in Eq.~(\ref{mellin}) correspond to resummation of DL contributions to all orders in $\alpha_s$ whereas
$C_{NS}^{DGLAP}$ and $h_{NS}^{DGLAP}$ contain DL contributions of several first orders in $\alpha_s$ only. On the other hand, they contain terms
important at moderate $x$, which should be accounted for.
This can be done in such a way: \\
\textbf{(i)} Subtract DL contribution from $C_{NS}^{DGLAP}$ and $h_{NS}^{DGLAP}$. We denote  $\widetilde{C}_{NS}^{DGLAP}$ and $\widetilde{h}_{NS}^{DGLAP}$
the result of such subtraction. \\
\textbf{(ii)} Add $\widetilde{C}_{NS}^{DGLAP}$ and $\widetilde{h}_{NS}^{DGLAP}$ to the DL expressions $C_{NS}$ and $h_{NS}$ respectively. We denote the result
$\widetilde{C}_{NS}^{DGLAP}$ and $\widetilde{h}_{NS}^{DGLAP}$:

\begin{eqnarray}\label{chtilde}
\widetilde{C}_{NS} &=&  C_{NS} + \widetilde{C}^{DGLAP}_{NS},
\\ \nonumber
\widetilde{h}_{NS} &=&  h_{NS} + \widetilde{h}^{DGLAP}_{NS}.
\end{eqnarray}

Replacing $C_{NS}$ and $h_{NS}$ in Eq.~(\ref{melling1}) by $\widetilde{C}_{NS}$ and $\widetilde{h}_{NS}$, we obtain the interpolation formula for $g_1^{NS}$ valid at
any $x$. The subtraction \textbf{(i)}  is necessary for avoiding  the double counting. Finally, $\alpha_s$
in the $\omega$-space
should be
replaced by the couplings
$A^{\prime}$ and $A$.  In relation to $h_{NS}$ it means replacing
$\alpha_s$ in the factor $2 \alpha_s/\pi$ by $A(\omega)$ and multiplying
$f^{(+)}_8$ by $A^{\prime}(\omega)/A (\omega)$,
see Ref.~\cite{egtg1a,egtg1b,egtg1c} for detail\footnote{Similar replacements should also be done in
expressions for $f^{(+)}_8$ }. After these replacements have been done, we arrive at the formulae for $g_1^{NS}$ valid at
arbitrary $x$ and accounting for the running coupling effects at the same time.
Generalization of the singlet $g_1$ on the case of arbitrary
$x$ and accounting for the running coupling effects can be done exactly the same way.

The small-$x$ asymptotics of $g_1^{NS}$ is of the Regge type. It is
obtained by applying the Saddle-Point method to Eq.~(\ref{melling1}), so the intercept is the rightmost singularity of the integrand.
The same features are true for the asymptotics $\bar{g}_1$ of the singlet. It is given by the following expression:

\begin{equation}\label{as}
\bar{g}_1 = \frac{\kappa}{\ln^{3/2}(1/x)} x^{- \Delta}\left(\frac{Q^2}{\mu^2}\right)^{\Delta/2}
= \frac{\kappa}{\ln^{3/2}(1/x)}\left(\frac{Q^2}{x^2 \mu^2}\right)^{\Delta/2},
\end{equation}
where $\kappa$ is a numerical factor and $\Delta$ is a general notation for the intercept. The intercepts were calculated for several interesting cases:

\textbf{Case A: }
$\alpha_s$ is fixed. We denote the intercept $\Delta_{fix}$. It was obtained in Ref.~\cite{bers}  is ( $N = 3$)

\begin{equation}\label{deltafix}
\Delta_{fix} = 3.45(\alpha_s N/2\pi)^{1/2}.
\end{equation}

\textbf{Case B: }
$\alpha_s$ is fixed and quark contributions are dropped. We denote the intercept $\Delta_g$. It obtained in Ref.~\cite{bers}
 and  confirmed in Refs.~\cite{kovch2}-\cite{smalldis}
 is

\begin{equation}\label{deltag}
\Delta_{g}= 3.66 (\alpha_s N/2\pi)^{1/2}.
\end{equation}

\textbf{Case C: }
$\alpha_s$ is running in each vertex of every Feynman graph involved according to Eqs.~(\ref{asmallx}, \ref{asmallxpi}). We
denote the intercept $\omega_0$ calculated in Ref.~\cite{egtg1b}  is

\begin{equation}\label{deltarun}
\omega_0= 0.86.
\end{equation}

Numerical value of
$\omega_0$ depends on $\mu$. The value $\omega_0 = 0.86$ was fixed by applying Principle of Minimal Sensitivity
suggested in  Ref.~\cite{pms}.
It corresponds to setting  $\mu$ at the scale

\begin{equation}\label{muscale}
\mu \approx 10 \Lambda_{QCD}.
\end{equation}

 Such a way of the scale setting is similar to the one applied in Ref.~\cite{kim} to
 the BFKL Pomeron.
It is interesting to note that the value of $\omega_0$ in Eq.~(\ref{deltarun})  remarkably agrees with the estimate

\begin{equation}\label{omegaprime}
\omega^{\prime}_0 = 0.88 \pm 0.14
\end{equation}
obtained in Ref.~\cite{koch} by
 extrapolating the HERA date to the region of $x \to 0$. The small difference between $\omega_0$ and  $\omega^{\prime}_0$ can be
attributed to the impact of  non-perturbative contributions and sub-leading perturbative contributions as well.

Estimating the applicability region of the small-$x$ asymptotics was considered in Ref.~\cite{egtg1c}.
To this end, the ratio $R_{as} = \bar{g}_1/g_1$
 was plotted against $x$ in Fig.~\ref{helfigRa}.
\begin{figure}\label{helfigRa}
\includegraphics[width=.4\textwidth]{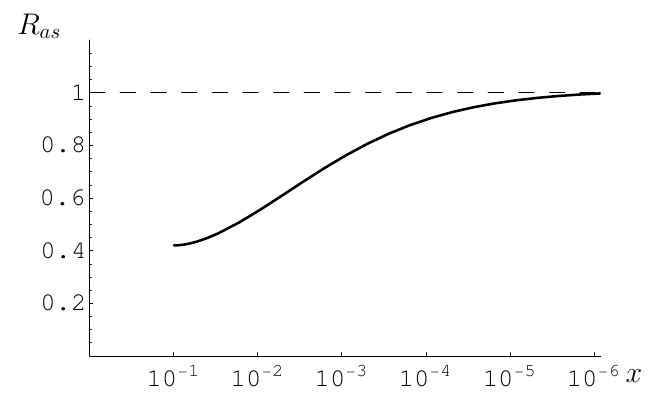}
\caption{\label{helfigRa} Dependence of $R_{as}$ on $x$ at small $Q^2$}
\end{figure}
The plot demonstrates that $R_{as} \approx 0.9$ at $x \approx x_0 = 10^{-6}$, so
 the small-$x$ asymptotics of $g_1$ can reliably represent $g_1$ at

 \begin{equation}\label{asreg}
x \leq x_0 = 10^{-6},
 \end{equation}
which we address as the  applicability region of the small-$x$ asymptotics. $R_{as}$ becomes small outside the region (\ref{asreg}). For instance,
the asymptotics of $g_1$ is almost twice less than $g_1$ at $x \approx 10^{-3}$. However, in the approaches where the total summation of leading
logarithmic contributions is mimicked by the factors $\sim x^{-a}$ in the fits (see Eq.~(\ref{fit})), the value of $a$ is always kept
greater than the genuine intercept in order to increase the impact of such factors at $x > x_0$.

\section{Evolution of helicity at small $x$}
\label{sec:sectii}

In the present Section, we consider evolving the quark and gluon helicities with respect to $x$. As we stated it in Sect.~1, we leave detailed investigating the $Q^2$-evolution
of the helicities for the future. Here we just notice that our preliminary estimates show that the impact of the $Q^2$-evolution on our conclusions  in the interval $1 < Q^2 < 10$~GeV$^2$
are small.

\subsection{Notations and definitions}
\label{sec:sectiia}

For the sake of convenience,
we will use throughout the paper the following notations instead of $\Delta \Sigma (x)$ and $\Delta G (x)$:

\begin{equation}\label{deltasghqg}
\Delta \Sigma (x) \equiv h_q(x),~~~\Delta G (x) \equiv h_g(x)
\end{equation}
  and define

\begin{eqnarray}\label{hqgdef}
H_q(a, b) &=& \int_{a}^{b} dx h_q (x),
\\ \nonumber
H_g(a,b) &=& \int_{a}^{b} dx h_g(x),
\end{eqnarray}
where $h_q$ and $h_g$ are evolved helicity distributions for quarks and gluons respectively. Throughout the paper we will address
$H_{q,g}(a, b) $ as the integral helicities.
So, the quark and gluon spins in terms of the integral helicities are:

\begin{eqnarray}\label{sqghtot}
S_q &=& \frac{1}{2} H_q(0,1),
\\ \nonumber
S_g &=& H_g(0,1).
\end{eqnarray}

Then express $\bar{S}_q$ and $\bar{S}_g$ of Eq.~(\ref{sexp}) through the integral helicities:

\begin{eqnarray}\label{sbar}
 \bar{S}_q &=& \frac{1}{2} H_q (x_1,1),
\\ \nonumber
\bar{S}_g &=& H_g (x_2,1)
\end{eqnarray}
and similarly define $S^{\prime}_q $ and $S^{\prime}_g $:

\begin{eqnarray}\label{sprime}
S^{\prime}_q &=&   \frac{1}{2} H_q (0,x_1),
\\ \nonumber
S^{\prime}_g &=&  H_g (0,x_2),
\end{eqnarray}
with $x_{1,2}$ corresponding to Eqs.~(\ref{x1},\ref{x2}) and

\begin{eqnarray}\label{hqgzero}
H_q(\emph{}x_1,1)  &=& \int_{x_1}^{1} dx h_q (x),
\\ \nonumber
H_g(x_2,1)  &=& \int_{x_2}^1 dx h_g (x),
\\ \nonumber
H_q(0,x_1)  &=& \int_{0}^{x_1} dx h_q (x),
\\ \nonumber
H_g(0,x_2)  &=& \int_{0}^{x_2} dx h_g (x).
\end{eqnarray}

Combining Eqs.~(\ref{sbar}) and (\ref{sprime}), obtain

\begin{eqnarray}\label{shqg}
S_q &=&  \bar{S}_q + S^{\prime}_q,
\\ \nonumber
S_g &=& \bar{S}_g +  S^{\prime}_g.
\end{eqnarray}

So, the forthcoming problem is calculating $H_q (0,x_1)$ and $H_g (0,x_2)$.

\subsection{Straightforward   calculation of  $S_q$ and $S_g$ }
\label{sec:sectiib}

Expressions for $h_q$ and $h_g$ contain contributions of large and small momenta of virtual partons, so they cannot be calculated in
the framework of Perturbative QCD. As usual, we use the QCD Factorization concept.
As is well-known, QCD Factorization represents $h_q$ and $h_g$ as convolutions of the perturbative parton-parton amplitudes $f_{ij}$
(where $i,j = q,g$)  and non-perturbative parton distributions $\Phi_{q,g}$:

\begin{eqnarray}\label{fact}
h_q &=& f_{qq} \otimes \Phi_q + f_{qg} \otimes \Phi_g,
\\ \nonumber
h_g &=& f_{gq} \otimes \Phi_q + f_{gg} \otimes \Phi_g.
\end{eqnarray}

Eq.~(\ref{fact}) is written in a general form, so it holds for any kind of QCD Factorization, though the
convolution symbol $\otimes$ means different sets of integrations, depending on the kind of Factorization. For the sake of simplicity,  we apply Collinear Factorization in the present paper but it is not a necessary restriction.
In Eq.~(\ref{fact}), $f_{ij}$  are the longitudinal spin-flip amplitudes of the forward parton-parton scattering. They were obtained in
 Ref.~\cite{egtg1b} (see also the overview \cite{egtg1c}) 
and they are valid at arbitrary $x$.  Generally speaking, both $h_{q,g}$ and $f_{ij}$ (i,j = q,g) depend on $x$ and $Q^2$. However, the $Q^2$-dynamics is not vital for our goal because the RHIC data
were taken at fixed $Q^2$, namely $Q^2 \approx 10$ GeV$^2$.
 So, in the present paper we neglect the $Q^2$-dependence of $h_{q,g}$ and $f_{ik}$ and will use the expressions for the
helicities obtained in
Ref.~\cite{egtg1b}. They include the total resummation of DL contributions and at the same time
account for the running QCD coupling effects. On the other hand, we admit that the $Q^2$ dynamics should be investigated  and we
are going to account for it in the future.

In contrast to $f_{ij}$, the
initial spin-dependent  parton distributions $\Phi_{q,g}$ are constructed on basis of phenomenological
considerations.
It was shown in Ref.~\cite{egtg1c} and briefly reproduced in Sect.~IIB  that $\Phi_{q,g}$ at small $x$ in the framework of Collinear Factorization
can be approximated  by constants which we denote $N_q$ and $N_g$. Distributions $\Phi_{q,g}$ (and $N_{q,g}$) are
of essentially non-perturbative origin.
Because of that they cannot be calculated with regular QCD means, so they are fixed from experiment. We also
follow this strategy and express
the constants $N_{q,g}$ through the experimental data given by
Eq.~(\ref{sexp}).
Combining Eqs.~(\ref{hqgdef},\ref{sbar}) and (\ref{fact})
, we obtain

\begin{eqnarray}\label{hqgnqg}
\bar{S}_q &=&  \frac{N_q}{2} \int^1_{x_1} dx f_{qq} (x) +  \frac{N_g}{2} \int^{1}_{x_1} dx f_{qg} (x),
\\ \nonumber
\bar{S}_g &=& N_q \int^1_{x_2} dx f_{gq} (x) + N_g \int^{1}_{x_2} dx f_{gg} (x).
\end{eqnarray}

Numerical values of $\bar{S}_{q,g}$ are known from  Eq.~(\ref{sexp}). So, performing integrations in (\ref{hqgnqg}), we arrive at the system of algebraic equations for $N_q,N_g$. Solving it, we specify
$N_q$ and $N_g$. Then, combining (\ref{hqgzero}) and (\ref{fact}), obtain

\begin{eqnarray}\label{hqgzeroh}
2S^{\prime}_q   &=&  H_q(0,x_1) =  N_q \int_{0}^{x_1} dx f_{qq} (x) +  N_g \int_{0}^{x_1} dx f_{qg} (x),
\\ \nonumber
S^{\prime}_g   &=& H_g(0,x_2) =  N_q \int_{0}^{x_2} dx  f_{gg} (x) + N_g \int_{0}^{x_2} dx f_{gg} (x).
\end{eqnarray}

All ingredients in the rhs of the both equations in (\ref{hqgzeroh}) are already known, so performing integrations over $x$, we obtain
$H_q(0,x_1)$ and $H_g(0,x_2)$. Substituting them in (\ref{shqg}) allows to specify $S_q$ and $S_g$. \\

Expressions for all $f_{ij}$ look similar to each other, though do not coincide, see Refs.~\cite{egtg1b,egtg1c}.
Despite that difference, $g_1$ and $f_{ij}$  manifest identical small-$x$ asymptotics of the Regge type (cf. Eq.~(\ref{as})):

\begin{equation}\label{g1as}
g_1 \sim f_{ij} \sim \xi^{-3/2}~ x^{- \omega_0} = \xi^{-3/2}~ e^{\omega_0 \xi},
\end{equation}
with $\xi = \ln \left(1/x\right)$ and $\omega_0$ being the intercept. When the running coupling effects are taken into account,
$\omega_0$ can be found with numerical calculations only. Its value is given by Eq.~(\ref{deltarun}).
The
 Regge asymptotics of parton distributions, structure functions and other interesting objects are given by simple expressions like Eq.~(\ref{g1as}),
 and because of that they are often used instead of their parent amplitudes. However,
we once more worn that they should not have been used outside their applicability region $x \leq x_0$, with $x_0 = 10^{-6}$, see Sect.~2.1.

\section{Estimating  $S_q$ and $S_g$}
\label{sec:sectiii}

The straightforward way to estimate $S_{q,g}$ is applying Eqs.~(\ref{hqgnqg},\ref{hqgzeroh}). They operate with the integral helicities
$H_{q,g}(a,b)$ mostly at small $x$, where DL terms dominate over other contributions.
However,
the helicity amplitudes $f_{q,g}$  are given by quite complicated expressions (see Refs.~\cite{egtg1b,egtg1c}) in the $\omega$-space.
It makes technically difficult applying them, so we would like to approximate $H_{q,g}(a,b)$ by simpler expressions $\widetilde{H}_{q,g}(a,b)$
 having maximal resemblance with $H_{q,g}(a,b)$.
The integration  region in (\ref{hqgnqg},\ref{hqgzeroh}) is pretty far from the applicability region of the asymptotics,
so $f_{q,g}$  cannot be approximated by their asymptotics.
We suggest below a simple approximation for $f_{ij}$, which has the correct asymptotics (\ref{g1as}) and on the other hand  is
pretty close to  $f_{ij}$ within the integration region in (\ref{hqgnqg},\ref{hqgzeroh}). In other words, we
construct an interpolation
formula for the quark and gluon helicities.

\subsection{Approximation of  $f_{ij}$}
\label{sec:sectiiia}

The starting point is the expression (see Refs.~\cite{egtg1b,egtg1c}) for amplitude $M_{gg}$ of the elastic $2 \to 2$ -scattering of gluons in the forward kinematics in the ladder approximation,
with all virtual partons being gluons.
We write it in terms of the Mellin transform:

\begin{equation}\label{mellin}
M_{gg} = \int_{-\imath \infty + c}^{\imath \infty + c} \frac{d \omega}{2 \pi \imath} x^{- \omega} F_{gg} (\omega),
\end{equation}
where the Mellin amplitude $F_{gg} (\omega)$ in DLA is

\begin{equation}\label{fgg}
F_{gg} (a,\omega) = 4 \pi^2 \left[\omega - \sqrt{\omega^2 - a}\right],
\end{equation}
with $a = 4 \alpha_s N/\pi$. Integration over $\omega$ in Eq.~(\ref{mellin}) runs along the $\Im \omega$ axis to the right of the rightmost singularity $\omega_0 = \sqrt{a}$.
Let us notice that replacement of the ladder gluons by quarks results into replacement of $a$ by $2\alpha_s C_F/\pi$, with $C_F = 4/3$. Integrating Eq.~(\ref{mellin})
over $\omega$ yields

\begin{equation}\label{agg}
M_{gg} (a,\xi) = - 4 \pi \frac{\sqrt{a}}{\xi} I_1 \left(\xi \sqrt{a}\right),
\end{equation}
where $I_1$ denotes the modified Bessel function.
In order to obtain the imaginary part of $M_{gg}$, we should recover correct analytic properties of $M_{gg}$, i.e. replace $\xi$ by  $\xi - \imath \pi$ and expand
$M_{gg} (a,\xi - \imath \pi)$ in series:

\begin{equation}\label{mser}
M_{gg} (a, \xi - \imath \pi) \approx M_gg(a,\xi) - \imath \pi \frac{d M_{gg} (a,\xi)}{d \xi}.
\end{equation}
As a result we obtain

\begin{equation}\label{imm}
\Im M_{gg}(a,\xi) = - \pi \frac{d M_{gg} (a,\xi)}{d \xi} = 4 \pi^2 \sqrt{a} \frac{d }{d \xi} \left( \frac{I_1 \left(\xi \sqrt{a}\right)}{\xi}\right)
=  4 \pi^2 \sqrt{a} ~\frac{ I_2 \left(\xi \sqrt{a}\right)}{\xi} .
\end{equation}

The small-$x$ asymptotics of the Bessel functions is well-known:

\begin{equation}\label{besselas}
I_{\nu} (\xi \sqrt{a})/\xi \sim  \xi^{-3/2} e^{\xi \sqrt{a}}
\end{equation}
at any $\nu$, so
the intercept of $\Im M_{gg}$ is $\sqrt{a}$. The simplest way to include  the impact of quark DL
contributions on $\Im M_{gg}$  and account for the running coupling effects at the same time is
replacing $a$ in Eq.~(\ref{imm}) by $\omega_0$ of Eq.~(\ref{deltarun}), obtaining thereby  $\Im M_{gg}(\omega_0,\xi)$
which
has the correct small-$x$ asymptotics
 (\ref{g1as}).  On the other hand, using the power expansion of $I_2  \left(\xi \sqrt{\omega_0}\right)/\xi$
 demonstrates that   $\Im M_{gg}(\omega_0,\xi)$
approximates the perturbative content of $f_{ij}$ much better  than its small-$x$ asymptotics
$e^{\xi \sqrt{\omega_0}}$.
Now let us get busy with specifying non-perturbative contributions to the helicities (cf. Eq.~(\ref{hqgnqg})).
In order to fix them  we write
our approximation formulae for the helicities $\widetilde{h}_{q,g}(x) $ as follows:

\begin{eqnarray}\label{hqgmod}
\widetilde{h}_q (x) = C^{\prime}_q  \frac{I_2 \left(\xi \sqrt{\omega_0}\right)}{\xi},
\\ \nonumber
\widetilde{h}_g (x) = C^{\prime}_g \frac{I_2 \left(\xi \sqrt{\omega_0}\right)}{\xi},
\end{eqnarray}
where $C^{\prime}_{q.g}$ are arbitrary factors. They accommodate both perturbative and non-perturbative factors. Therefore,
Eq.~(\ref{hqgdef}) takes the following form:

\begin{eqnarray}\label{tildehab}
\widetilde{H}_q(a, b) &=& C^{\prime}_q \int_{a}^{b} dx \frac{ I_2 \left(\xi \sqrt{a}\right)}{\xi} = C_q  \int_{z_b}^{z_a} dz e^{ - z/\omega_0} \frac{ I_2 \left(z\right)}{z},
\\ \nonumber
\widetilde{H}_g(a,b) &=& C^{\prime}_g \int_{a}^{b} dx  \frac{ I_2 \left(\xi \sqrt{a}\right)}{\xi} = C_g \int_{z_b}^{z_a} dz e^{ - z/\omega_0} \frac{ I_2 \left(z\right)}{z},
\end{eqnarray}
with $z = \omega_0 \xi$ and $C_{q,g} = \omega_0 C^{\prime}_{q,g}$ are arbitrary factors.
It is convenient to represent both $\widetilde{H}_q (x_1,1), \widetilde{H}_g (x_2,1)$ and $\widetilde{H}_q (0,x_1),\widetilde{H}_g (0,x_2)$
corresponding to Eq.~(\ref{hqgzero}) as follows:

\begin{eqnarray}\label{hcab}
\widetilde{H}_q (x_1,1)  &=& C_q A_q,
\\ \nonumber
\widetilde{H}_g (x_2,1)
&=&
 C_g A_g,
\\ \nonumber
\widetilde{H}_q (0,x_1)  &=& C_q B_q,
\\ \nonumber
\widetilde{H}_g (0,x_2)
&=&
 C_g B_g,
\end{eqnarray}
where

\begin{eqnarray}\label{abz}
A_q   &=&   \int_0^{z_1} dz e^{- z/\omega_0}   \frac{I_2(z)}{z},
\\ \nonumber
A_g  &=&   \int_0^{z_2} dz e^{- z/\omega_0} \frac{I_2(z)}{z},
\\ \nonumber
B_q &=&    \int^{\infty}_{z_2} dz e^{- z/\omega_0} \frac{I_2(z)}{z},
\\ \nonumber
B_g  &=&   \int^{\infty}_{z_2} dz e^{- z/\omega_0} \frac{I_2(z)}{z}.
\end{eqnarray}
with
\begin{equation}\label{z12}
z_1 = \omega_0 \ln (1/x_1),~~z_2 = \omega_0 \ln (1/x_2).
\end{equation}

 We specify $C_{q,g}$ , using Eq.~(\ref{sexp}):

\begin{eqnarray}\label{cqgz}
C_q  &=&  2 \bar{S}_q /A_q,
\\ \nonumber
C_g  &=&  \bar{S}_g /A_g
\end{eqnarray}

After $C_{q,g}$ have been specified, we can estimate $H_q (0,x_1)$ and $H_g (0,x_2)$:

\begin{eqnarray}\label{hbqg}
\widetilde{H}_q (0,x_1)  &=& C_q B_q = 2 \bar{S}_q  \left( B_q/A_q\right) ,
\\ \nonumber
\widetilde{H}_q (0,x_2)  &=& C_g B_g = \bar{S}_g \left( B_g/A_g\right).
\end{eqnarray}

Therefore,

\begin{eqnarray}\label{sqsgab}
S_q &=& \bar{S}_q \left(1 + B_q/A_q\right),
\\ \nonumber
S_g  &=&   \bar{S}_g \left( 1 + B_g/A_g\right).
\end{eqnarray}

\subsection{Numerical calculations}
\label{sec:sectiiib}


Substituting in Eq.~(\ref{abz}) numerical values

\begin{equation}\label{lambdaz12}
\omega_0 = 0.86,~~z_1 = \omega_0 \ln \left(1/ 10^{-3}\right) = 5.94,~~z_2 = \omega_0 \ln \left(1/ (5.10^{-2})\right)=2.57,
\end{equation}
we obtain

\begin{eqnarray}\label{abnum}
A_q  &=&  0.138,
\\ \nonumber
A_g  &=&  0.0874,
\\ \nonumber
B_q  &=&  0.0243,
\\ \nonumber
B_g  &=&  0.0747.
\end{eqnarray}

Combining (\ref{abnum}) with (\ref{sqsgab}) leads to the following estimates for the evolved intervals $\Delta S_q, S_g$:

\begin{eqnarray}\label{sqgab}
\Delta S_q  &=&   \Delta  \bar{S}_q \left( 1 + 0.0243/0.138\right) = 1.18~\bar{S}_q
\\ \nonumber
\Delta S_g  &=&    \Delta \bar{S}_g \left( 1 + 0.0747/0.0874\right) = 1.85 ~\bar{S}_g.
\end{eqnarray}

Using Eq.~(\ref{sexp}) for numerical estimates for $\bar{S}_{q,g}$, obtain

\begin{eqnarray}\label{sqgtot}
0.18 \leq S_q \leq 0.24,
\\ \nonumber
0.24 \leq S_g \leq 0.72.
\end{eqnarray}
Adding up them, we conclude that the proton spin is within the following interval:

\begin{equation}\label{sp}
0.42 \leq S_P \leq 0.72,
\end{equation}
which includes the value $S1/2$.  This is illustrated in Fig.~\ref{helfig3}, where the projections of $\Delta S_q$ and $\Delta S_g$ on the
line $S_q + S_g = 1/2$ partly overlap. All $S_{q,g}$ from the overlapping the projections of $S_{q,g}$ overlap and therefore
 the values of $S_{q,g}$ within the intervals $\Delta S_q$ and $\Delta S_g$ correspond to $S_P = 1/2$.

\begin{figure}\label{helfig3}
\includegraphics[width=.6\textwidth]{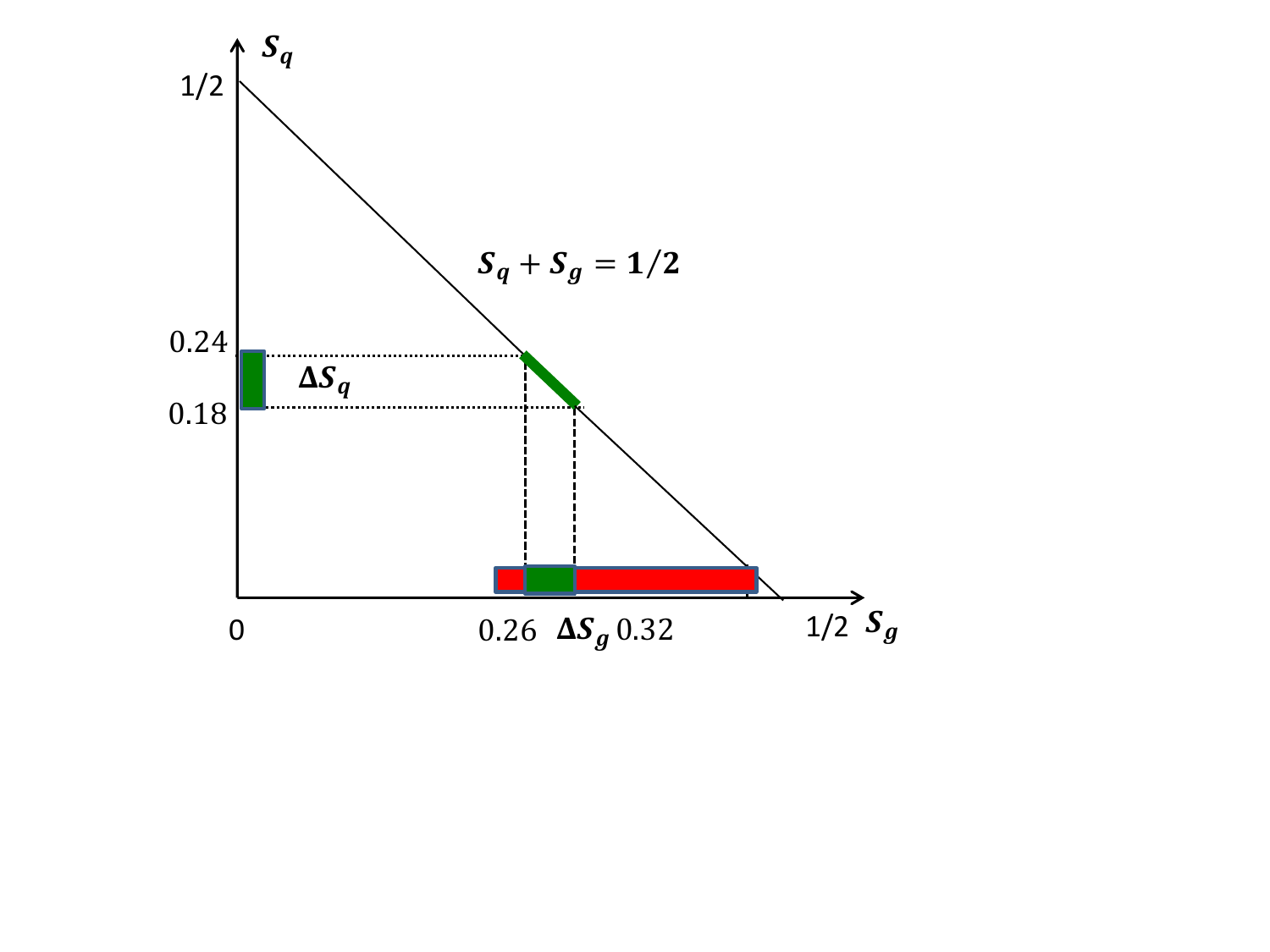}
\caption{\label{helfig3} Projections of $\Delta S_q$ and $\Delta S_g$ on the line $S_q + S_g = 1/2$ overlap, i.e. the
values of $S_{q,g}$ of Eq.~(\ref{sqgtot}) within these intervals (green blocks) obey the requirement $S_q + S_g = 1/2$.}
\end{figure}

\section{Summary and outlook}
\label{sec:sectiv}

We have demonstrated that using DLA for calculation of the parton
contributions $S_q$ and  $S_g$ to the proton spin $S_P$ ensures perfect agreement  with
the value $S_P =1/2$ (see Eq.~(\ref{sp}) and Fig.~3).
$S_{q,g}$  are defined as integrals of the parton helicities, each of them includes both perturbative components (which
we calculated in DLA) and non-perturbative components which we specified with making use of the RHIC data.  We presented
the straightforward approach to description of the proton spin based on the results obtained in Refs.~\cite{egtg1a,egtg1b,egtg1c}.
Then we
suggested the shortcut by constructing the convenient approximation for the quark and gluon helicities at small $x$.
As a result, we obtained
in Eq.~(\ref{sqgtot}) the estimates
of the parton helicities, which led to agreement with Eq.~(\ref{spdef}).

It is stated in Refs.~\cite{w1,w2} that contributions from very small $x$ are unessential for solving the Proton Spin Problem whereas
our opinion is opposite. This contradiction is easy to explain: the DGLAP fits (see Eq.~(\ref{fit})) are constructed in such a way as to increase the impact of large and medium
$x$, where the perturbative components of DGLAP work well, and diminish the impact of small $x$, where the lack of resummation of the DL terms
makes the DGLAP expressions for coefficient functions inadequate.
Indeed, the
factor $x^{-a}$ introduced in the DGLAP fits to guarantee the fast growth at small $s$  contains the parameter $a$ which is chosen greater than the genuine intercept $\omega_0$ (see Ref.~\cite{egtg1c} for detail). A similar sense has introducing
the term $cx^d$.

Unlike  Refs.~~\cite{agr}-\cite{global}
operating with the small-$x$
asymptotics  of the helicities both inside and outside their applicability region,
we use explicit DL
expressions for the parton helicities. This allows us to adequately treat
the region, where  $x$ are small, but not asymptotically small, without getting a help from the fits.

In contrast to Refs.~\cite{agr}-\cite{w2},  we  did not consider  contributions of the quark and gluon Angular Orbital Momenta to the proton spin and do not trace the
$Q^2$- dynamics of $S_{q,g}$ because our goal was to verify whether the sum of the RHIC data
and
contributions $S^{\prime}_{q,g}$, was compatible with the requirement $S_P = 1/2$ requirement. Next our step will be to consider the $Q^2$-dependence.
 Then. we admit that studying parton AOMs is an important issue and plan to get busy with it.  To begin with, we plan to check out whether accounting for
 AOMs is compatible with using Collinear Factorization or  $k_T$-Factorization should be used in this case.

\acknowledgments

We are grateful to Yu.V.~Kovchegov and S.M.~Osipov for useful communications.

\appendix

\section{Relations between helicities and the structure function $g_1$ in DLA}

QCD factorization represents the singlet structure function $g_1$ through the convolutions of perturbative components $g_1^{(q,g)}$ and initial parton distributions $\Phi_{q,g}$:

\begin{equation}\label{g1fact}
g_1  = g_1^{(q)} \otimes \Phi_q +  g_1^{(g)} \otimes \Phi_g.
\end{equation}

For the sake of simplicity we consider below $g_1$ at $Q^2 \approx \mu^2$, so that $g_1 = g_1 (x)$. More general expressions for $g_1$ in
the kinematics $Q^2 > \mu^2$ can be found in Ref.~\cite{egtg1c}. It is convenient to represent $g_1^{(q,g)}$ in
terms of the Mellin transform:

\begin{equation}\label{melling1s}
g_1^{(q,g)} (x) = \frac{e^2_q}{2} \int_{- \imath \infty}^{\imath
\infty} \frac{d \omega}{2 \pi \imath} x^{- \omega}\;\omega\;
F_{q,g}(\omega)\; \Phi (\omega).
\end{equation}

Mellin amplitudes $F_{q,g}$ obey the following IREEs:

\begin{eqnarray} \label{fh}
\omega  F_q &=&  c_q + F_q f_{qq} + F_g f_{gq},
  \\ \nonumber
\omega F_g &=&  F_q f_{qg} + F_g f_{gg},
\end{eqnarray}
where $c_q$ is the Born contribution. Amplitudes $f_{ij}$ are defined in Eq.~(\ref{fact}). They are
perturbative components of the parton helicities $h_{q,g}$. Explicit expressions for $c_q$ and $f_{ik}$ can be found in Ref.~\cite{egtg1c}.
Solution to Eq.~(\ref{fh}) is

\begin{eqnarray} \label{fhsol}
  F_q &=&    c_q \frac{\omega - f_{gg}}{\omega^2 - \omega (f_{qq} + f_{gg})- (f_{qq}f_{gg} - f_{qg}f_{gq}) },
  \\ \nonumber
F_g &=&  c_q \frac{f_{qg}}{\omega^2 - \omega (f_{qq} + f_{gg})- (f_{qq}f_{gg} - f_{qg}f_{gq}) }.
\end{eqnarray}

According to the Saddle-Point method, small-$x$ asymptotics $\bar{g}_1 (x)$ of $g_1(x)$ is   $\bar{g}_1 (x) \sim x^{- \Delta}$, with the intercept $\Delta$ being
the largest root (the stationary point) of the equations

\begin{equation}\label{aseq}
\ln (1/x) + \frac{d}{d \omega} \ln F_{q,g} = 0
\end{equation}
 at $x \to 0$. In other words, $\Delta$ is the rightmost singularity of $F_{q,g}$. Both $F_q$ and $F_g$ are made out of $f_{ik}$,
 so their rightmost singularities coincide with the ones of $f_{ij}$. It is the reason why the intercepts of the small-$x$ asymptotics of $g_1$
 and the parton helicities are the same.

\end{document}